resub2.tex

\documentstyle[aps,twocolumn,floats]{revtex}
\begin{document}
\newcommand\1{$\spadesuit$}
\newcommand\2{$\clubsuit$}
\def\be{\begin{equation}}
\def\ee{\end{equation}}
\def\ba{\begin{eqnarray}}
\def\ea{\end{eqnarray}}
\tighten
\draft
\twocolumn[\hsize\textwidth\columnwidth\hsize\csname 
@twocolumnfalse\endcsname
\title{On the Generation of a Scale-Invariant Spectrum of Adiabatic
Fluctuations in Cosmological Models with a Contracting Phase}

\author{Fabio Finelli$^{1,2)}$ and Robert Brandenberger$^{1,3)}$}
\address{$^{1)}$ Theory Division, CERN, CH-1211 Gen\`eve 23, Switzerland}
\address{$^{2)}$ I.A.S.F. - Sezione di Bologna, C.N.R., Via Gobetti 101,
40129 Bologna, Italy.\\ e-mail: finelli@tesre.bo.cnr.it}
\address{$^{3)}$
 Department of Physics, Brown University, Providence, RI 02912, USA.\\
e-mail: rhb@het.brown.edu}

\date{February 15, 2002}
\maketitle

\begin{abstract}

In Pre-Big-Bang and in Ekpyrotic Cosmology, perturbations
on cosmological scales today are generated from quantum
vacuum fluctuations during a phase
when the Universe is contracting (viewed in the Einstein
frame). The backgrounds studied to date do not yield
a scale invariant spectrum of adiabatic fluctuations.
Here, we present a
new contracting background model (neither of Pre-Big-Bang
nor of the Ekpyrotic form) involving a single
scalar field coupled to gravity in which a scale-invariant
spectrum of curvature fluctuations and gravitational waves results. The
equation of state of this scalar field corresponds
to cold matter. We demonstrate that if this contracting phase can be matched
via a nonsingular bounce to an expanding Friedmann cosmology,
the scale-invariance of the curvature
fluctuations is maintained. We also find new
background solutions for Pre-Big-Bang and for Ekpyrotic
cosmology, which involve two scalar fields with
exponential potentials with background values which are evolving in time. 
We comment on the difficulty of obtaining a scale-invariant
spectrum of adiabatic fluctuations with background solutions
which have been studied in the past.
 
\end{abstract}

\vspace*{1cm}
]

\section{Introduction}

Both Pre-Big-Bang and Ekpyrotic cosmology are attempts to
construct alternatives to inflationary cosmology by
introducing ideas of string theory to cosmology. The
Pre-Big-Bang scenario \cite{Gasperini:1993em} is based on considering the
dilaton at an equal footing to the gravitational field,
motivated by the fact that these are the important low energy 
degrees of freedom. The action is given (in the string frame) by
\begin{equation} \label{PBBaction}
S \, = \, {1 \over {2 \kappa^2}} \int d^4x \sqrt{-g} e^{- \varphi}
\bigl( R - (\partial \varphi)^2 \bigr) \, ,
\end{equation}
where $\varphi$ denotes the dilaton and $\kappa^2 \equiv M_{\rm pl}^{-2}
\equiv 8 \pi G$. The Universe is assumed
to begin in a cold empty state with an accelerating dilaton.
The initial evolution is dominated by the effects of the dilaton,
and in the string frame yields super-exponential expansion. 
In the Einstein frame, this corresponds to a contracting phase
with a scale factor $a(t)$ given by $a(t) \sim (-t)^{1/3}$ (the
time $t$ is negative in this phase). By
a duality transformation, this solution is related to an
expanding Friedmann-Robertson-Walker cosmology. Without corrections
to the action (\ref{PBBaction}), however, the initial 
dilaton-dominated (pre BB) branch and the late time 
expanding (post BB) branch are separated by a singularity.
Since the Hubble radius decreases faster than the physical
wavelength corresponding to fixed comoving scales, quantum
fluctuations on microscopic scales during the pre BB branch can
be stretched to scales which are cosmological at the present time,
as in the case of inflationary cosmology. See e.g. 
\cite{Veneziano:2000pz,Lidsey:1999mc} for recent reviews of Pre-Big-Bang
cosmology.

The Ekpyrotic scenario \cite{Khoury:2001wf} (see also \cite{Khoury:2001bz})
assumes that the visible Universe is a boundary
brane in five dimensional bulk space-time, and that the heating event which
corresponds to the Big Bang of Standard cosmology resulted
from the collision of this brane with a parallel one which is
attracted to it by an inter-brane potential $V(\varphi)$. The
dynamics is described by a four-dimensional toy model
in which the separation of the branes in the extra dimension
is modeled as a scalar field $\varphi$. The effective action is taken
to be
\begin{equation} \label{EKPaction}
S \, = \,  \int d^4x \sqrt{-g}
\bigl({1 \over {2 \kappa^2}} R 
+ {1 \over 2}(\partial \varphi)^2 - V(\varphi) \bigr) \, ,
\end{equation}
with a potential which for values of $\varphi$ relevant to
the generation of cosmological fluctuations is given by
\begin{equation} \label{pot}
V(\varphi) \, = \, - V_0 e^{- \sqrt{2 \over p}(m_p)^{-1} \varphi} \, ,
\end{equation}
where $0 < p \ll 1$ and $m_p$ denotes the 4-d Planck mass  
(using the notation of \cite{Lyth:2001pf}). The branes
are assumed to start out widely separated and at rest.
In this case, the energy is negative and the scale factor
associated with the action (\ref{EKPaction}) is contracting with
$a(t) \sim (-t)^{p}$ (the time $t$ is again negative in this phase).
The time $t = 0$ corresponds to a
singularity of the four dimensional model (\ref{EKPaction}),
as in the case of Pre-Big-Bang cosmology \footnote{See, however,
\cite{Kallosh:2001ai,Kallosh:2001du,Enqvist:2001zk,Rasanen:2001hf} 
for criticism of the scenario.}. As in the case of
Pre-Big-Bang cosmology, comoving scales contract less fast
than the Hubble radius during this phase, and thus it is again
possible that microscopic sub-Hubble scale fluctuations during the
phase of contraction produce perturbations on cosmological
scales today.

Neither for Pre-Big-Bang cosmology nor in the Ekpyrotic scenario 
\footnote{Note that the fluctuation generation mechanism in the
recently proposed cyclic model \cite{Steinhardt:2001st} is the
same as in the Ekpyrotic scenario, and hence also does not yield
a scale-invariant spectrum of fluctuations.}
is a scale-invariant spectrum of adiabatic fluctuations generated
at the level of the single field actions described above. A heuristic
way to understand this is to note that the initial values of the
fluctuations when they exit the Hubble radius are set by the Hubble
constant. The Hubble constant is increasing rapidly as a function of time
in both scenarios, and thus a deeply blue spectrum of initial fluctuations
will result (spectral index $n = 4$ in the case of Pre-Big-Bang cosmology,
$n = 3$ in the case of the Ekpyrotic scenario). Careful studies
taking into account the gravitational dynamics on super-Hubble scales
confirmed this result both for Pre-Big-Bang cosmology \cite{Brustein:1995kn} 
and in the Ekpyrotic scenario 
\cite{Lyth:2001pf,Brandenberger:2001bs,Hwang:2001ga,Lyth:2001nv,Tsujikawa:2001ad}
\footnote{In the case of the Ekpyrotic scenario,
the calculation of \cite{Khoury:2001zk} yields a different result, but at
least in our opinion is flawed because it is based on an ad hoc matching
condition at the bounce which (as already mentioned in 
\cite{Brandenberger:2001bs}) does not yield the correct result when applied
to power-law inflation (see also 
\cite{Martin:2001ue} and \cite{Durrer:2001qk}
for recent work on this issue.)}. 

The idea of obtaining the ``big bang'' of our Universe from a
previous phase of cosmological contraction is, however, very
interesting. In Section 2 we discuss a model consisting of
scalar field matter with an equation of state $P = 0$, $P$ denoting
pressure, obtained with the value $p=2/3$ for the exponential potential
in Eq. (\ref{pot}) (we note that for $p=2/3$ the potential is positive,
i.e. $V_0 < 0$, as opposed to the case of the Ekpyrotic model 
\cite{Khoury:2001wf}, where the potential is negative, i.e. $V_0 > 0$). 
In this background, the quantum vacuum fluctuations of the
field during the phase of contraction, matched to an
expanding Friedmann cosmology at a nonsingular bounce, yield a scale-invariant
spectrum of curvature fluctuations \footnote{Note that we are
assuming the absence of any initial classical fluctuations, as is
done in the Ekpyrotic scenario.}. Such a model is obtained here by 
considering an exponential potential for the scalar field. In the PBB 
scenario exponential potentials for the dilaton may be generated by 
non-perturbative effects or by considering non-critical string theory (a 
cosmological constant in the string frame can generate an exponential 
potential in the Einstein frame).

In order to connect the contracting phase to an expanding phase, it
is necessary to assume that at sufficiently high curvatures corrections
to Einstein gravity become important, yielding a nonsingular bounce.
Similar ideas are invoked to achieve a graceful exit in Pre-Big-Bang
cosmology. In Section 3 we apply matching conditions 
\cite{Hwang:1991an,Deruelle:1995kd} corresponding
to continuity of the induced metric and of the extrinsic curvature
for the infrared modes to calculate the induced curvature
fluctuations in the expanding phase. We find that the dominant mode
of the curvature perturbation $\zeta$ in the expanding phase
inherits the scale-invariance of the growing mode of the
contracting phase. A new aspect of this matching problem is that
the dominant mode of $\zeta$ increases on super-Hubble scales in
the contracting phase, in contrast to what occurs in inflationary cosmology,
where it is constant.
    
In Section 4 we study backgrounds with two scalar matter fields.
Note that both for Pre-Big-Bang cosmology and in the Ekpyrotic
scenario, there are other light fields which should be included in
the respective actions (\ref{PBBaction}) and (\ref{EKPaction}). In both
cases there are axion and moduli fields which could play an important
role. These fields can be dynamical during the collapse phase 
\cite{Copeland:vi}.

In the case of Pre-Big-Bang cosmology, it was realized in 
\cite{Copeland:1997ug} that, in the presence of moving extra dimensions, 
axion fluctuations are amplified, and that the motion of
the extra dimensions can be chosen such that a scale 
invariant spectrum of isocurvature perturbations results
(for more work 
along these lines see e.g. \cite{Durrer:1998sv,Melchiorri:1999km}).
However, such a primordial spectrum of inhomogeneities 
seeded by axion fluctuations is ruled out
by the latest CMB anisotropy results 
(see e.g. \cite{Melchiorri:1999km} for an analysis of this issue).

In Section 4 of this paper, 
we present a new class of two scalar field backgrounds. 
These backgrounds have an interpretation both in the case of the
Pre-Big-Bang scenario and in Ekpyrotic cosmology. In the former case they
correspond to a dilaton-axion background with exponential dilaton potential,
in the latter case they involve the scalar field $\varphi$ representing
the separation of the two branes and a rolling axion. 

\section{Fluctuations in Single Field Background Models}

We begin with a brief review of the analysis of the
spectrum of cosmological fluctuations in single field background 
models.

To linear order in fluctuations (and neglecting gravitational
waves and vector modes), the metric can be written as (see e.g.
\cite{Mukhanov:1992me} for a comprehensive review)
\begin{equation} \label{longitudinal}
ds^2 \, = \, a^2(\eta)\bigl[(1 + 2\Phi)d\eta^2 - (1 - 2\Phi)dx^idx_i \bigr]
\, ,
\end{equation}
where $\eta$ is conformal time. We have used the fact that
if matter consists of scalar fields there is to linear order no 
anisotropic stress. 

Via the Einstein constraint equations,
the linear gravitational fluctuations (described by $\Phi$) are
coupled to the matter field fluctuations. In the single matter
field case, with matter fluctuations denoted by $\delta \varphi$,
a convenient and gauge-invariant variable is \cite{Mukhanov:1988jd}
(see also \cite{Lukash})
\begin{equation} \label{variables}
v \, =  \, a \bigl( \delta \varphi + {{{\dot \varphi}} \over H} \Phi
\bigr) \equiv a Q \, , 
\end{equation}
with an overdot denoting the derivative with respect to physical time $t$.
In particular, in the action for joint metric and matter fluctuations,
$v$ is a canonically normalized field, and hence it is useful to
quantize the fluctuations in terms of it. We will be studying the
linear perturbation equations for this variable in momentum space,
with $k$ standing for comoving momentum.
 
In an expanding Universe described by the Einstein equations, 
a convenient variable to use to track the amplitude of the fluctuations
on super-Hubble scales is $\zeta$, the curvature perturbation
in comoving gauge 
\cite{Bardeen:1980kt,Lyth:1985gv}. In variable $\zeta$ is related to
the metric fluctuations via
\begin{equation}
\zeta \, = \, {2 \over 3} {{\Phi + H^{-1}{\dot \Phi}} \over {1 + w}}
+ \Phi \, .
\end{equation}
and it is related to $Q$ and $v$ as follows:
\be
\zeta = \frac{H}{\dot \phi} Q = {v \over z} \, ,
\label{zetaQ}
\ee
with
\begin{equation}
z \, = \, a {{\dot{\varphi}} \over H} \, .
\end{equation}

In an expanding Universe and in the absence of entropy fluctuations, 
$\zeta$ is constant on scales larger than the Hubble radius, as can be
seen from its equation of motion which is
\begin{equation} \label{zetaeq}
{\dot{\zeta}} \, = - {H \over {\dot{H}}} {{k^2} \over {a^2}} \Phi \, .
\end{equation} 
In inflationary cosmology, it is thus useful to calculate the
magnitude of $\zeta$ at the time when the fluctuation scale becomes
larger than the Hubble radius, to use the constancy of $\zeta$ to evolve
until the time when the scale re-enters the Hubble radius,
and to infer the values of $\Phi$ and ${\dot \Phi}$ (which determine,
for example, the spectrum of CMB anisotropies) at that time.

However, in the case of a contracting Universe one must (even in
the absence of entropy fluctuations) be more careful, since the
term in (\ref{zetaeq}) proportional to $k^2$ may grow. 
In the case of Pre-Big-Bang cosmology, the growth is only logarithmic
in $\eta$, and for the potential used in the original version of
the Ekpyrotic model there is
no growth of $\zeta$ at all. Hence, in these models 
$\zeta$ remains a good variable
to follow the magnitude of the density fluctuations on super-Hubble
scales.

In the Einstein frame, 
the equations of motion for cosmological perturbations reduce to the
following equation for the Fourier mode of the variable $v$ defined 
in (\ref{variables}) with comoving wavenumber $k$ (we suppress the
index $k$ on $v$)
\begin{equation}
v^{''} + \bigl(k^2 - {{z^{''}} \over z}\bigr)v \, = 0 \, ,
\end{equation}
where a prime denotes the derivative with respect to conformal time $\eta$.

Making use of the
background equations of motion, this equation becomes
\begin{eqnarray} \label{basiceom}
v^{''} + \bigl(&k^2& - {{a^{''}} \over a} + a^2 V^{''} \\
&+& 2 a^2 \bigl[{{{\dot H}} \over H} + 3H \bigr]^{\cdot} \bigr) v = 0
\, , \nonumber
\end{eqnarray}
where $V$ is the potential of the matter field $\varphi$.
From this equation is is clear that on scales much smaller than the 
Hubble radius, $v$ is oscillating with frequency given by $k$. The
vacuum state normalization of $v$ on these scales is
\begin{equation}
v \, = {1 \over {\sqrt{2 k}}} e^{-i k \eta} \, .
\end{equation}

We consider backgrounds which correspond in the Einstein frame
to power law contraction
\begin{equation}
a(t) \, \propto \, (-t)^p \, .
\end{equation} 
In this case, the last two terms within the parentheses 
multiplying $v$ in (\ref{basiceom}) cancel, and the
equation reduces to
\begin{equation} \label{veq}
v^{''} + \bigl(k^2 - {{p(2p - 1)} \over {(p-1)^2}}{1 \over {\eta^2}}\bigr)v 
\, = \, 0 \, .
\end{equation}
Its solution can be expressed in terms
of Bessel functions $Z_{\nu}$:
\begin{equation} \label{vsol}
v \, = \, \sqrt{-\eta} Z_{|\nu|}(-k\eta)
\end{equation} 
where the index $\nu$ is related to the index $p$ by
\begin{equation}
{\nu^2 - 1\over 4} \, = \, {{p(2p -1)} \over {(p - 1)^2}} \, 
\label{nuindex}
\end{equation}
and therefore 
\begin{equation} 
\nu = \frac{1}{2} \frac{1 - 3p}{1-p} \,.
\end{equation}
As can be seen from the large argument expansion of the Bessel
functions, this solution automatically has the correct vacuum normalization.

Making use of the long wavelength limit of the Bessel functions, and
of the fact that for our class of backgrounds ${\dot{\varphi}} / H$ is
independent of time,
we obtain the following power spectrum of the variable $\zeta$:
\begin{equation}
P_{\zeta}(k) \, = \, {{k^3} \over {2 \pi^2}} |\zeta|^2 \,
 \sim  \, p k^{3 - 2\nu} \, . 
\end{equation}
Thus, to obtain a scale-invariant spectrum of adiabatic
fluctuations, we require $\nu = 3/2$, whereas in the
single field Ekpyrotic scenario with $p \sim 0$ one obtains $\nu \sim 1/2$ and
thus spectral index $n \sim 3$. In the single field
Pre-Big-Bang scenario the resulting values are $p = 1/3$, $\nu = 0$
and therefore $n = 4$.

An interesting background is obtained if $p = 2/3$. In this
case, $\nu = 3/2$ and hence a scale-invariant spectrum of adiabatic curvature
fluctuations is generated in the collapsing phase \footnote{This
result was already noted in the work of \cite{Wands:1998yp}, and more  
recently in \cite{Brandenberger:2001bs} (it can be also 
seen from Fig. 3 of \cite{Tsujikawa:2001ad}). In the earlier
work \cite{Wands:1998yp}, however, the transition to an expanding phase
was not discussed.}. This background
corresponds to a contracting Universe dominated by cold matter with
equation of state $P = 0$ (with $P$ denoting pressure). The cold
matter in this case is modeled by a scalar field with an exponential
potential. Note that the fluctuations in this model behave differently
than the fluctuations in a hydrodynamical model with $P = c_s^2 = 0$,
where $c_s^2$ is the speed of sound. This can be seen by comparing
the equation of motion (\ref{basiceom}) for scalar-field-induced
fluctuations with the corresponding equation for hydrodynamical matter
(see Chapter 5 of \cite{Mukhanov:1992me}). Note that modeling cold
matter by a scalar field with nontrivial potential will avoid the
Jeans instability problem on small scales which plagues $P = 0$
hydrodynamical matter. Note also that such a contraction ($a(t) \sim 
(-t)^{2/3}$) solve the horizon problem. 

\section{Spectrum of Fluctuations After the Bounce}

Let us consider the contracting model with $p = 2/3$ introduced above
and imagine that it is connected via a nonsingular bounce to an
expanding Friedmann-Robertson-Walker cosmology. In the following
we will argue that in such a model the scale-invariant spectrum of
$\zeta$ connects through the bounce to a late time scale-invariant
spectrum in the expanding phase. Thus, our model yields an
alternative to cosmological inflation in providing a mechanism for
producing a scale-invariant spectrum of adiabatic fluctuations on
scales which could, provided that the phase of contraction lasts
sufficiently long, be of cosmological interest today.

The graceful exit
problem faced in order to obtain such a bounce is similar to what
is required to obtain a graceful exit in Pre-Big-Bang cosmology. In
the latter case, there are indications that higher curvature
and string loop effects can yield a nonsingular bounce 
\cite{Gasperini:1996fu,Maggiore:1998cz,Foffa:1999dv,Brustein:1997cv,Cartier:1999vk} (see also \cite{Antoniadis:1993jc,Easther:1996yd}. Back-reaction
effects also can play an important role \cite{Ghosh:1999xm}.
There is also a construction 
\cite{Brandenberger:1998zs,Easson:1999xw} based on a Lagrangian with
higher order corrections of the nonsingular Universe construction 
\cite{Mukhanov:1991zn,Brandenberger:1993ef} which yields such a
regular bounce. Since the corrections to Einstein gravity will be
important only very close to the bounce, and since we are interested
in scales which at the bounce are much larger than the Hubble radius,
it appears reasonable to model the bounce (for the purpose of matching
the pre-bounce and post-bounce fluctuations) as a gluing of two
Einstein Universes at a fixed surface specified by some physical
criterion (see, however, the concerns raised in 
\cite{Lyth:2001nv,Martin:2001ue} on this issue).

The naive expectation is that the dominant mode of the
curvature fluctuation $\zeta$ after the bounce (which is constant in
time) is given by the dominant mode prior to the bounce, the mode
which in our background has a scale-invariant spectrum. This
is what happens in inflationary cosmology with reheating modeled as
a discontinuous change in the equation of state
\cite{Bardeen:1983qw,Brandenberger:1984tg},
Pre-Big-Bang cosmology \cite{Deruelle:1995kd} and
in the Ekpyrotic scenario with a nonsingular bounce 
\cite{Brandenberger:2001bs}. Note however, that a similar ``plausibility''
argument applied to the variable $\Phi$ fails in both of the above cases.
In both Pre-Big-Bang and Ekpyrotic cosmology the pre-bounce growing mode
of $\Phi$ does not contribute to leading order in the wavenumber $k$ to
the dominant post-bounce mode of $\Phi$. Thus, in order to be
able to draw conclusions about the post-bounce spectrum, it is
necessary to carefully match the fluctuation variables at the bounce. 

The general relativistic matching conditions demand that at the
boundary surface the induced surface metric and the extrinsic
curvature be continuous \cite{Hwang:1991an,Deruelle:1995kd}. As
matching surface one can choose either a constant energy
density surface (from the point of view of
longitudinal gauge) or a constant scalar field surface \footnote{See
\cite{Martin:2001ue} for a justification of this choice of
the matching surface.}. For super-Hubble
scale fluctuations, the difference between these two surfaces is
of order $k^2$ and does not effect the results to leading order in $k^2$.
The matching on a constant energy density surface implies the
continuity of $\zeta$ and $\Phi$ across the surface. In
Pre-Big-Bang cosmology \cite{Deruelle:1995kd} and in the Ekpyrotic
scenario \cite{Brandenberger:2001bs} this leads to the conclusion that
the growing mode of $\Phi$ during the collapse phase (which in the
case of the Ekpyrotic scenario has a scale-invariant spectrum)
does not couple to leading order in $k^2$ to the dominant (constant)
mode of the post-bounce phase. In contrast, the late time value
of $\zeta$ is the same as at the bounce. These results are
confirmed in studies of cosmological perturbations in generalized
Einstein theories \cite{Tsujikawa:2001ad,Starobinsky:2001xq,Cartier:2001is} 
which yield a bounce (see also \cite{Hwang:1999gf,hwangnoh}).

At the end of Section II we have shown that, for our new
background, the growing mode
of $\zeta$ in the contracting phase obtains a scale-invariant
spectrum. We need to show that to leading order in $k^2$ there is
a non-vanishing coupling between the pre-bounce growing mode of
$\zeta$ and the post-bounce dominant mode. Note that the rapid growth of
$\zeta$  is very different from what occurs in
the inflationary Universe, in Pre-Big-Bang cosmology and in the
Ekpyrotic scenario, and that the analysis of the matching conditions
needs to be reconsidered. 

We begin with the following general solution (to leading order in $k^2$)
for $\zeta$ (see e.g. Eq. (12.27) in \cite{Mukhanov:1992me}):
\be
\zeta \, = \, D \, + \, S \int{{{d\eta} \over {z^2}}} \, ,
\ee
where $D$ and $S$ are constant coefficients.
In our background, the S mode is the growing one, whereas in
inflationary cosmology it is decaying and thus subdominant. Making use
of (\ref{zetaeq}), we can find the corresponding form of $\Phi$:
\be
\Phi \, = \, - {1 \over {2 m_p^2}} {{\cal{H}} \over {a^2}} 
\bigl({S \over {k^2}}\bigr) \, ,
\ee
where ${\cal H}$ is the Hubble constant in conformal time. This gives
the mode of $\Phi$ which is decaying in inflationary cosmology. It
also shows that this ``decaying'' mode of $\Phi$ effects the value
of $\zeta$ to order $k^2$ (to leading order it cancels out). There
is also a constant mode of $\Phi$ which determines the constant
mode (D mode) of $\zeta$. The coefficients of the constant modes of
$\Phi$ and $\zeta$ are related by a function of the equation of
state of the background which contains no k-dependence. Thus, in order
to perform the matching analysis we consider the following forms for
$\Phi$ and $\zeta$:
\begin{eqnarray}
\zeta \, &=& \, D \, + \, S f(\eta) \, , \\
\Phi \, &=& \, \alpha D \, + \, \beta \frac{S}{k^2} {{\cal{H}} \over 
{a^2}} \, , \nonumber
\end{eqnarray}
where $f(\eta) = \int{{{d\eta} \over {z^2}}}$ and where the coefficients
$\alpha$ and $\beta$ depend only on the equation of state.

It is now straightforward to calculate the consequences of the continuous
matching of $\Phi$ and $\zeta$ across the bounce for the coefficients
of the two modes of $\zeta$ before and after the bounce. Quantities before
the bounce will be denoted by a superscript $-$, those after the
bounce by a superscript $+$. Simple algebra yields
\begin{eqnarray}
D^+\bigl(1 - {{\alpha^+ f^+ a^2 k^2} \over {\beta^+ \cal{H}^+}}\bigr) \, &=& \,
D^-\bigl(1 - {{\alpha^- f^+ a^2 k^2} \over {\beta^+ \cal{H}^+}}\bigr) 
\nonumber \\
& & + \, S^-\bigl(f^- - {{\beta^-} \over {\beta^+}}
{{\cal{H}^-} \over {\cal{H}^+}} f^+\bigr) \, .
\end{eqnarray}

In the Ekpyrotic scenario, it follows from (\ref{vsol}) and from the fact that
$\nu = 1/2$ that $D^- \sim k^{-1/2}$ and $S^- \sim k^{1/2}$. Hence, the
D-mode of $\zeta$ after the bounce has spectral index $n = 3$, in
agreement with our earlier matching results \cite{Brandenberger:2001bs}.
However, for our present background we have $\nu = 3/2$ and hence 
$D^- \sim k^{3/2}$ and $S^- \sim k^{-3/2}$. Since the matching of the
$S^-$ mode to the $D^+$ mode is not suppressed by factors of $k^2$, the
post-bounce $D^+$ mode inherits the scale-invariance of the 
pre-bounce $S^-$ mode. Thus, we have shown that our background yields
a scale-invariant spectrum of fluctuations at late times. 

Let us now return to the analysis of Section II and
determine the amplitude of the scale-invariant 
spectrum of curvature perturbations at (and, as we have shown above, 
therefore also after) the bounce in our model with $p = 2/3$. 
As follows from Eq. (\ref{zetaQ}) and from the background values of
${\dot \varphi}$ and $H$:
\begin{equation}
\zeta = \frac{H}{\dot \varphi} Q = \sqrt{\frac{p}{2}} \frac{Q}{M_{\rm pl}}
\label{spectrum}
\end{equation}
According to (\ref{vsol}), the normalized solution for $Q$ is:
\be
Q = e^{[i (\tilde{\nu} + 1/2) \pi/2]} \frac{\sqrt{- \pi \eta}}{2 a}
H_{\tilde{\nu}} (- k \eta)  
\ee
where $H$ denotes the Hankel function, $\tilde{\nu} = |\nu|$, 
and where the scale factor is 
expressed in terms of conformal time as 
\be \label{sfactor}
a(\eta) = ( -(1-p) M_{\rm{pl}} \eta)^{\frac{p}{1-p}} \, .
\ee
By inserting the long-wavelength limit 
($-k \eta \ll 1$) of the Hankel function one gets:
\begin{equation} 
P_\zeta(k) = \frac{k^3}{2 \pi^2} |\zeta|^2 = p 2^{2\tilde{\nu} - 3} 
\frac{\Gamma(\tilde\nu)^2}{2 \pi^3} 
\left[ \frac{{\cal H}(1-p)}{M_{\rm{pl}} p} \right]^{ 4 \tilde \nu} 
\frac{M_{\rm{pl}}^{2 \tilde{\nu} - 3}}{k^{2 \tilde \nu-3}} \,.
\end{equation}
For $p = 2/3$ one has an amplitude in agreement with observations for 
${\cal H}_*/M_{\rm{pl}} \sim 10^{-1}$, where ${\cal H}_*$ 
indicates the absolute value of the
Hubble rate at the bounce (when the contraction stops).

The model with $p = 2/3$ which we are considering also generates
a scale-invariant spectrum of gravitational waves, as already realized in
\cite{Starobinsky:ty} and \cite{Wands:1998yp}. 
This can be
seen immediately since ${\tilde{h}} = a h$, where $h$ is the
amplitude of the tensor perturbation of the metric, obeys the
same equation of motion as the scalar fluctuation variable $v$.
This is true in all models with an exponential potential for the scalar 
field \cite{stewlyth:1992}.
In order to compare the amplitudes of the
power spectra $P_{\zeta}$ for scalar metric fluctuations and
$P_{h}$ for gravitational waves
in the collapsing phase, we first need to divide $v$ by $M_{pl}$
(for dimensional reasons). Then, using the relation (\ref{zetaQ}) between
$\zeta$ and $v$ and the background values of $H$ and $\dot{\varphi}$
we immediately obtain
\be
{{P_{\zeta}} \over {P_h}} \, = {p \over 2} \, .
\ee
This is a definite prediction resulting from the analysis during
the contracting phase. 

\section{A Background with Two Evolving Fields}

We now turn to the question of whether it might be possible to generate a
scale-invariant spectrum of adiabatic fluctuations in Pre-Big-Bang
cosmology and in the Ekpyrotic scenario by turning on nontrivial
background time-dependence of a suitably chosen second scalar field.
As a first step, we will have to construct new background solutions.
We will consider the action
\begin{eqnarray} \label{newaction} 
S \, = \,  \int d^4x \sqrt{-g}
&\bigl(&{1 \over {2 \kappa^2}} R + {e^{\alpha \varphi/M_{\rm pl}} \over
2}(\partial
\sigma)^2
\\
& + & {1 \over 2}(\partial \varphi)^2 - V(\varphi) \bigr) \, , \nonumber
\end{eqnarray}
with an exponential potential for $\varphi$:
\be
V(\varphi) = - V_0 e^{-\beta \varphi/M_{\rm pl}} \, .
\label{potential_varphi}
\ee

In the Ekpyrotic scenario, the interpretation of this action is that
we add to the single field action (\ref{EKPaction}) 
(with $\varphi$ representing the separation of the branes) an axion
field $\sigma$ with non-minimal coupling to gravity. In Pre-Big-Bang
cosmology, the interpretation of (\ref{newaction}) is to add an
exponential potential for the dilaton $\varphi$ and an axion-like field
(non-minimally coupled in the Einstein frame) 
to the original action of (\ref{PBBaction}). For $\alpha=0$ the field 
$\sigma$ should be interpreted as a modulus field. 

We now look for self-consistent analytical solutions of the background
equations:
\begin{eqnarray}
{\cal H}^2 &=& \frac{1}{3 M^2_{\rm pl}} \left( \frac{\varphi^{'2}}{2} +
a^2 V(\varphi) +
e^{\alpha \varphi/M_{\rm pl}} \frac{\sigma^{'2}}{2} \right) \label{ham2}
\\
\varphi '' &+& 2 {\cal H} \varphi' + a^2 V_\varphi = \frac{\alpha}{2
M_{\rm pl}} 
e^{\alpha \varphi/M_{\rm pl}} \sigma^{' 2}  \label{bulkbrane} \\
\sigma'' &+& (2 {\cal H} + \alpha \frac{\varphi'}{M_{\rm pl}} ) \sigma'
= 0 \, . \label{axion}
\end{eqnarray}

We consider the following ansatz for $a(\eta)$ and $\varphi(\eta)$:  
\begin{eqnarray} \label{ansatz2}
a(\eta) \, &=& \,(- M_{\rm pl} (1-p) \eta)^\frac{p}{1-p}
\\
\varphi (\eta)\, &=& \, A \log (- M_{\rm pl} (1-p) \eta) \label{ansatz1}
\end{eqnarray}

Eq. (\ref{axion}) can be immediately integrated giving:
\be
\sigma ' = C \frac{e^{-\alpha \varphi/M_{\rm pl}}}{a^2}
\ee
where $C$ is an integration constant.
We have the following parameters: $\alpha, \beta, p, A, C$.
By imposing that all the terms in the equation for $\varphi$
(\ref{bulkbrane}) have the same time dependence we obtain (assuming $V_0$ 
and $C$ are nonzero):
\begin{eqnarray} \label{constr2}
\beta \frac{A}{M_{\rm pl}} - \frac{2p}{1-p} &=& 2 \\
\alpha \frac{A}{M_{\rm pl}} + \frac{4p}{1-p} &=& 2
\end{eqnarray}
and these relations lead to the constraint 
\be \label{constr}
\alpha/\beta \, = \, 1-3p \, .
\ee
Note that this relation implies that for the case of Pre-Big-Bang
cosmology with $p = 1/3$ 
our action describes a modulus field $\sigma$ minimally
coupled in the Einstein frame, whereas in the case of the Ekpyrotic
scenario, $\sigma$ corresponds to an axion.

From the equation for ${\cal H}'$ \footnote{It is not an independent
equation, but is obtained from the time derivative of the Hamiltonian
constraint (\ref{ham2}), making use of the equations of motion for 
$\varphi$ and $\sigma$. 
From this equation it is easy to derive another constraint
among the parameters.}:
\be
{\cal H}' - {\cal H}^2 = - \frac{1}{2 M_{\rm pl}^2} \left[ \varphi^{' 2} + 
\sigma^{' 2} e^{\alpha \varphi/M_{\rm pl}} \right] 
\ee
we get:
\be
\left( \frac{A}{M_{\rm pl}} \right)^2  + \left( \frac{C}{M^2_{\rm pl}(1 - p)}
\right)^2 = \frac{2 p}{(1-p)^2} \,.
\label{circle}
\ee
The interpretation of this result is that $A/M_{pl}$ and 
$C/[M_{pl}^2(1 - p)]$ are constrained to 
be on a circle of radius proportional to $\sqrt{p}$. From the energy 
constraint (\ref{ham2}) we obtain:
\be
- \frac{V_0}{M^4_{\rm pl}} = p (3p -1) \,. 
\label{v0}
\ee
This concludes all of the independent relations between the parameters and
integration constants for the exact solution for the background obtained
from the ansatz (\ref{ansatz1} \ref{ansatz2}). 

Let us compare the dependence of $a$ and $V(\varphi)$ on $p$ and on time
in our two field solution with the corresponding scalings in the single
field model of Section II. Comparing (\ref{ansatz2}) with (\ref{sfactor}) 
it follows
that the dependence of $a$ on $p$ and on time is the same. Inserting 
(\ref{ansatz1}) into (\ref{potential_varphi}), 
making use of (\ref{v0}), and comparing with the result in the
single field case (see e.g. Eq. (12) of \cite{Brandenberger:2001bs}) it
follows that the dependence of $V(\varphi)$ on $p$ and on time is also
identical.
This implies that the dependence on $p$ of the spectrum and amplitude of 
gravitational waves is the 
same in the single field and multi field models. Eq. (\ref{circle}) 
shows that the sum of the kinetic terms in the multi field model with the 
axion is the same of the single field model.

A final remark for the case $\alpha = 0$, in which $\sigma$ corresponds 
to a modulus field. 
The solution obeying the ansatz (\ref{ansatz1}) constructed in this
section has a meaning
for $\alpha=0$, only in the case $p=1/3$, i.e. vanishing potential for 
$\varphi$. Only in this case analytic solutions exist, since both 
$\varphi$ and $\sigma$ are massless, and the global equation of state is 
stiff matter (pressure density equal to energy density). In the case of an 
exponential potential as in Eq. (\ref{potential_varphi}), the modulus 
$\sigma$ dominates the energy density for early times if $p < 
1/3$ and for late times if $p >1/3$. 

\section{Discussion}

In this paper we have discussed ways of obtaining a scale-invariant
spectrum of adiabatic fluctuations in models in which a contracting
Universe is matched via a nonsingular bounce to an expanding
Friedmann-Robertson-Walker cosmology. We have assumed that new
physics at high curvatures leads to a short period in which the
background evolution is not described by the Einstein equations,
thus enabling a transition from a contracting to an expanding phase
(from the point of view of the Einstein frame scale factor). 

Our first result is that a contracting Universe dominated by cold
matter modeled as a scalar field with an exponential potential
with appropriately chosen index can yield a scale-invariant spectrum
of curvature fluctuations (fluctuations in the variable $\zeta$)
in the contracting phase, which is matched at the bounce to a
scale-invariant spectrum during the expanding phase. 

Our second result is that it is possible in the context of Ekpyrotic
cosmology (and also of other models with a phase of power law
contraction with power $p \neq 1/3$) to obtain new background solutions 
by considering
a model with two scalar fields $\varphi$ and $\sigma$, with
exponential potential for $\varphi$ and non-minimal coupling for $\sigma$,
provided that the coefficients in the exponents of the potential
and describing the non-minimal coupling satisfy a particular relation.
In the context of Ekpyrotic cosmology, this corresponds to adding to
the usual four-dimensional effective action which contains a scalar
field $\varphi$ describing the separation of the branes an axionic
field $\sigma$. It is important that both fields have time-dependent
backgrounds. In the context of the Pre-Big-Bang scenario, our model
can be interpreted as adding to the dilaton-gravity action an
axion field, and assuming that there is an exponential potential for
the dilaton. 

Adding the new degree of freedom via the second scalar field leads to an
isocurvature mode of the fluctuations. In the PBB scenario, it was shown
\cite{Copeland:1997ug} that for suitable field backgrounds this
isocurvature mode has a scale invariant spectrum. However, because of the
absence of a potential, the isocurvature and adiabatic modes are not
coupled and a spectrum seeded by isocurvature axion fluctuations seems in 
conflict with observations \cite{Melchiorri:1999km}. In our backgrounds,
adiabatic and
isocurvature fluctuations are coupled and it might be possible 
to induce directly a scale invariant spectrum of adiabatic fluctuations.

This mechanism is an alternative to the one proposed recently 
based on axion decay to obtain a
scale-invariant spectrum of adiabatic curvature
fluctuations \cite{Enqvist:2001zp,moroi,Lyth:2001nq}.
One starts with a scale-invariant spectrum of isocurvature fluctuations
(like in the
work of \cite{Copeland:1997ug}), and assumes that the axion field which is
responsible for the scale-invariant spectrum decays at some late time
when it dominates the background energy density.

\vspace{0.5cm} 
\centerline{\bf Acknowledgments}
\vspace{0.2cm}

We wish to thank D. Wands for useful correspondence.
This research was supported in 
part by the U.S. Department of Energy under Contract 
DE-FG02-91ER40688, TASK A (R.B., Brown Univ.).  
Both authors thank the TH Division
at CERN for hospitality during the course of this work.




\end{document}